\documentclass[%
 reprint,
 amsmath,amssymb,
 aps,
 prl
]{revtex4-2}
\usepackage{lipsum}
\usepackage{diagbox}
\usepackage{url}
\usepackage{multirow}
\usepackage{xcolor}
\definecolor{dblue}{RGB}{25,25,125}
\usepackage[%
  colorlinks = true,
  urlcolor= dblue,
  linkcolor= dblue,
  citecolor= dblue
]{hyperref}
\usepackage{graphicx}
\usepackage{dcolumn}
\usepackage{bm}
\newcommand{\D}{\text{d}}

\begin{document}

\title{Detecting cosmological gravitational waves background after removal of compact binary coalescences in future gravitational wave detectors}

\author{Haowen Zhong}
\affiliation{%
 School of Physics and Astronomy, University of Minnesota, Minneapolis, MN 55455, USA
}%
\author{Rich Ormiston}%
 \author{Vuk Mandic}
\affiliation{%
 School of Physics and Astronomy, University of Minnesota, Minneapolis, MN 55455, USA
}%
\date{\today}
\begin{abstract}
The improved sensitivity of third generation gravitational wave detectors opens the possibility of detecting the primordial cosmological stochastic gravitational wave background (SGWB). Detection of the cosmological SGWB is facing a novel challenge: it will likely be masked by the foreground generated by a large number of coalescences of compact binary systems consisting of black holes and/or neutron stars. In this paper, we investigate the possibility of reducing this foreground by removing (notching) the individually resolved compact binary signals in time-frequency space. We establish that such an approach could be used to reach the SGWB sensitivity floor defined by the unresolved part of the compact binaries foreground, which we find to be between $\Omega_{\rm GW} \sim (9.1 \times10^{-12} - 8.6\times10^{-11})$ for a frequency independent energy density spectrum
and depending on the rate of coalescing binary neutron star systems. Since third-generation gravitational wave detectors will not be able to resolve all compact binaries, the unresolvable component of the compact binaries foreground may limit the SGWB searches with these detectors. 
\end{abstract}

\maketitle

\textbf{Introduction:} Since the first detection of a binary black hole (BBH) merger~\cite{GW150914} by the Advanced LIGO (aLIGO) \cite{aLIGO} and Advanced Virgo (aVirgo) \cite{aVirgo} detectors, the field of gravitational wave (GW) astrophysics has flourished. The first binary neutron star (BNS) merger detection~\cite{GW170817} was accompanied by a short Gamma Ray Burst detected by Fermi and INTEGRAL satellites~\cite{GW170817GRB} and then followed by the kilonova signal that evolved in ensuing days and weeks~\cite{GW170817MMA}. These discoveries have enabled a series of novel studies, including unprecedented tests of  General Relativity~\cite{GW170817GRB,PhysRevD.103.122002}, measurements of the Hubble constant~\cite{GW170817H0,Abbott_2021_H0}, studies of the origin and evolution of neutron stars \cite{GW170817progenitor,Abbott_2021_Pop}, and others.

As the sensitivity of GW detectors improves, one of the prime targets of future GW searches will be the stochastic gravitational wave background (SGWB), which arises as an incoherent superposition of many GWs generated by independent sources. SGWB can be of cosmological origin, and therefore detection of cosmological SGWB could reveal valuable information about the conditions of the early universe via the GW phenomenology of cosmic strings~\cite{CS1,CS2,CS3,CS4,CS5}, phase transitions~\cite{PT1,PT2,PT3,PT4,PT5}, primordial black holes~\cite{PBH1,PBH2,PBH3}, and others. The SGWB could also be of astrophysical origin~\cite{Regimbau_2011}---in particular, the background generated by compact binary coalescences (CBCs) of systems composed of black holes and/or neutron stars in distant galaxies would carry rich information about the population of compact binaries and about the formation of the large scale structure in the universe. 

Given the observed rate of individually detected BBH, BNS, and black-hole-neutron-star (BHNS) events, it is possible to estimate the energy density in GWs due to these systems integrated across the universe~\cite{O3stoch}. Under the assumption that these systems are all of stellar origin, the CBC foreground is likely to be stronger than the cosmological SGWB~\cite{Astrobg}. This therefore opens the question of whether the CBC foreground can be removed, thereby enabling a measurement of the cosmological SGWB. This question has been investigated in the past, including proposals of subtracting the CBC signals from the GW data~\cite{TaniaCE,subtractionSurabhi,CutlerHarms,ShwarmaHarms} and using global Bayesian schemes for estimating the SGWB along with all CBCs~\cite{Thrane_TBS,PhysRevLett.125.241101}. 

In this paper we investigate a simpler approach. Instead of subtracting or otherwise fitting for indivdual CBC signals, we investigate notching of the individually detected CBC signals in the time-frequency domain. This approach reduces the amount of available data to make the SGWB measurement, and is therefore less sensitive than the subtraction or Bayesian schemes. However, it is also less susceptible to biases arising from imperfect subtraction or statistical modeling. This approach is a relatively simple extension of the available infrastructure used for SGWB searches in LIGO and Virgo data. Hence, this approach serves as a reference point---it establishes the SGWB sensitivity achievable with future GW detectors using the existing SGWB search technology; future, more sophisticated techniques may be able to surpass this sensitivity.

We will investigate the efficacy of our approach by simulating realistic strain data for future GW detectors, including a population of compact binaries whose properties are drawn from the latest CBC catalog \cite{Abbott_2021_Pop}, and a truly stochastic GW background. We focus on the proposed third-generation detector Cosmic Explorer (CE) \cite{CE}. This detector will deploy similar technology to those of aLIGO and aVirgo, but its interferometer arms will be $10\times$ longer (40 km each), and the overall strain sensitivity will be $10\times$ (or more) better than that achieved by aLIGO/aVirgo. We assume that two such detectors will be operational, and to fully define this 2-detector network we assume the detectors to have the same locations and orientations of the current LIGO detectors at Livingston (L), LA and Hanford (H), WA.


\textbf{Stochastic GW background:} The normalized SGWB energy density spectrum is defined by
\begin{equation}
    \Omega_{\text{GW}}(f)=\frac{f}{\rho_{c,0}}\frac{\D \rho_{\text{GW}}(f)}{\D f}
\end{equation}
where $\D\rho_{\text{GW}}(f)$ is the GW energy density in the frequency band $(f,f+\D f)$, $\rho_{c,0}=3H_0^2c^2/(8\pi G)$ is the critical energy density required to close the universe, $c$ is the speed of light, $G$ is Newton's constant, and $H_0 = 67.9 {\rm \; km/s/Mpc}$ is the Hubble constant \cite{Hubble}.  In this work, the SGWB search will assume power-law SGWB of the form: 
\begin{equation}
    \Omega_{\rm GW}(f)=\Omega_\alpha\Big(\frac{f}{f_{\rm ref}}\Big)^\alpha
\end{equation}
where $\alpha$ is the power-law index of GW spectrum and $f_{\rm ref} = 25$ Hz is the reference frequency.

SGWB searches with terrestrial GW detectors rely on cross correlating strain data from two (or more) detectors. The strain time series detected by a single detector $h_{\text{I}}$ (I = H or L) is given by:
\begin{equation}
    h_{\text{I}}(t)=F^+_{\text{I}}h_+(t)+F^\times_{\text{I}}h_\times(t)
\end{equation}
where $F^+$ and $F^\times$ are antenna pattern functions corresponding to the plus mode ($h_+$) and the cross mode ($h_\times$) of gravitational waves~\cite{Antenna}. The time series is typically divided into segments of duration $T = 60-192$ sec, and data in each segment is Fourier transformed. In our application, we will need to track the time evolution of CBC signals, so we choose a shorter segment duration, $T = 4$ sec. The Fourier transforms in each segment are denoted by $\tilde{h}_{\text{I}}(t_i;f_j)$, where $t_i$ now denotes the time segment used to compute the Fourier transform and $f_j$ denotes a frequency bin. Following Refs. \cite{O3stoch,CC}, we  define the cross-correlation statistics $\hat{C}_{\text{IJ}}(t_i;f_j)$ 
and the corresponding variance estimator $\hat{\sigma}^2_{\text{IJ}}(t_i;f_j)$
for the baseline \text{IJ}:
\begin{eqnarray}
    \hat{C}_{\text{IJ}}(t_i;f_j) & = & \Big( \frac{20\pi^2f_j^3}{3H_0^2T} \Big) \frac{\Re[{\tilde{h}_{\rm{I}}^*(t_i;f_j)\tilde{h}_{\rm{J}}(t_i;f_j)}]} {\gamma_{\text{IJ}}(f_j)} \nonumber \\
    \hat{\sigma}_{\text{IJ}}^2(t_i;f_j) & = & \Big( \frac{20\pi^2f_j^3}{3H_0^2} \Big)^2 \frac{P_{n_{\rm{I}}}(t_i;f_j)P_{n_{\rm{J}}}(t_i;f_j)} {8T\Delta f\gamma_{\rm{IJ}}^2(f_j)}.
    \label{eq:stoch}
\end{eqnarray}
Here, $\gamma_{\text{IJ}}(f_j)$ is the overlap reduction function between two detectors \cite{CC}, $\Delta f = 0.25$ Hz is the frequency resolution of the Fourier transform, and $P_{n_{\text{I}}}(t_i;f_j)$ is the single-sided noise power spectral density (PSD) of the detector I for the time segment $t_i$. Past SGWB searches \cite{O3stoch} have operated in the weak-signal regime, which implies that the SGWB itself does not contribute significantly to the strain power of the detectors. These searches could then directly estimate the PSD from the strain data, typically using data from segments neighboring the segment $t_i$. In our study, due to the significant contribution of the CBC signals, we cannot assume the weak-signal regime and therefore cannot estimate the PSDs from the data. Instead, in our analysis we will use the same noise PSD that is used to simulate detector data. We recognize this as a possible challenge for the analysis of future real data, since the PSDs will not be known \textit{a priori}. 

After obtaining the statistic and its variance for each segment and each frequency bin, we combine the results so as to optimize the signal-to-noise:
\begin{eqnarray}
\hat{C}_{\text{IJ}} & = & \frac{\displaystyle{\sum_{ij} w(f_j) \hat{C}_{\text{IJ}}(t_i;f_j) \hat{\sigma}_{\text{IJ}}^{-2}(t_i;f_j)}} {\displaystyle{\sum_{ij} w^2(f_j) \hat{\sigma}_{\text{IJ}}^{-2}(t_i;f_j)}}\nonumber\\
\hat{\sigma}_{\text{IJ}}^{-2} & = & \sum_{ij} \frac{w^2(f_j)}{\hat{\sigma}^2_{\text{IJ}}(t_i;f_j)}.
\label{eq:combine}
\end{eqnarray}
Here, the weight $w(f_j)$ is defined by $w(f_j)=\Omega_{\text{GW}}(f_j)/\Omega_{\text{GW}}(f_{\text{ref}})$ \cite{O3stoch}. In our study we consider two cases: $\alpha=0$ and $\alpha=4$. With these definitions, $\hat{C}_{\text{\text{IJ}}}$ is the broadband detection statistic averaged over the entire observation data, normalized so that $\langle \hat{C}_{\text{\text{IJ}}} \rangle = \Omega_{\rm \alpha}$. 

\textbf{CBC Population:} A key aspect of our study is a realistic simulation of the CBC population's GW signals. To this end, we use the most recent information about the CBC population, extracted from individual CBC observations included in the third Gravitational-Waves Transient Catalog (GWTC-3) \cite{Population}. We include both BNS and BBH contributions; we do not include the BHNS contributions due to the relatively poorer current understanding of this population. 

For the BBH population we use the Power law-Peak (PP) model of the mass distribution  \cite{Population,parameters}. For the BNS events, we choose uniform distribution in mass between $M_\odot$ and $2M_\odot$ following \cite{Population,NSmass}. Angular parameters of the BBH and BNS models, including the cosine of the inclination angle, sky localization angles, and GW polarization are drawn from the uniform distribution. We also assume that the CBC component masses have no spin.

The redshift $z$ is drawn from the following probability distribution $p(z)$:
\begin{equation}
    p(z)=\frac{R_z(z)}{\displaystyle{\int_0^{10}}R_z(z)\D z},
\end{equation}
where $R_z(z)$ is the CBC merger rate per interval of redshift over the range $z\in[0,10]$ in the observer frame
\begin{equation}
    R_z(z)=\frac{R_m(z)}{1+z}\frac{\D V}{\D z}\Big|_z,
\end{equation}
$\D V/\D z$ is the comoving volume element, and $R_m(z)$ is the merger rate per comoving volume in the source frame:
\begin{equation}
    R_m(z)=\int_{t_{\min}}^{t_{\max}}R_f(z_f)P(t_d)\D t_d.
\end{equation}
Here, $R_f(z_f)$ is the binary formation rate as a function of the redshift at binary formation time $t_f=t_f(z_f)$ and $P(t_d)$ is the distribution of the time delay $t_d$ between the formation and merger. For BNS we set $t_{\min}=20$ Myr and for BBH we set $t_{\min}=50$ Myr; $t_{\max}$ is set to be the Hubble time for both cases. We model the time delay distribution $p(t_d) \propto 1/t_d$ and we model the binary formation rate as the star formation rate (SFR) whose expression is given by \cite{Redshift}:
\begin{equation}
    \text{SFR}(z)=\nu\frac{pe^{q(z-z_m)}}{p-q+qe^{p(z-z_m)}}
\end{equation}
with 
$z_m=2.00,p=2.37$ and $q=1.80$.

We limit the redshift range to $z\in[0,10]$, and we choose the normalization factors $\nu_{\text{BBH}}$ and $\nu_{\text{BNS}}$ so as to match the BBH and BNS merger rates to the observed ones \cite{Population}. More specifically, we choose $\nu_{\text{BBH}}$ so that $R_{\rm BBH}(z=0.2) = 28.1 {\rm \; Gpc^{-3} \; yr^{-1}}$. For BNS events, since the uncertainty in the local merger rate is significant, we consider three 
local merger rate values estimated by the  PDB(pair) Model, MS Model, and BGP Model \cite{Population}: $R_{\rm BNS} = 960 {\rm \; Gpc^{-3} \; yr^{-1}}$, $470 {\rm \; Gpc^{-3} \; yr^{-1}}$, and $99 {\rm \; Gpc^{-3} \; yr^{-1}}$, respectively. We refer to these three BBH+BNS models as Pop A, B and C. With these assumptions, we estimate there will be one CBC merger per $d = 3, 6$, and 24 seconds for these three different populations, respectively. 

\textbf{Mock Data Generation:} Our first step in generating mock strain data is to generate 
three 1-year long lists of CBC events corresponding to the Pop A, B, and C defined above. This is done using the \texttt{MDC\_Generation} package \cite{PhysRevD.92.063002,MDC}. We note that due to significant differences in CBC merger rates in the three populations, these three lists have significantly different numbers of CBC events.

We then use \texttt{Bilby} \cite{Bilby} to compute time-domain GW waveforms (TaylorT4 waveform model) for these three lists and add them to the strain noise time series generated assuming the expected noise PSD of the second stage Cosmic Explorer (CE2) \cite{CE}. The start time of each waveform is defined to be the moment when its chirp crosses 7 Hz frequency. Mock data is generated for two detectors (H and L), and then recorded in the standard frame file format. For some of the analyses discussed below we also add a mock SGWB signal, using the Matlab-based pipeline for SGWB search \cite{O3stoch,Stochastic}. 
This mock SGWB signal is chosen to have amplitude  $\Omega_\alpha = 4 \times 10^{-11}$, with $\alpha = 0$ or 4.

\textbf{Mock Data Analysis:}
The mock data frames are then processed using the standard Matlab-based pipeline for SGWB search \cite{O3stoch,Stochastic}. This pipeline performs the calculation defined in Eqs. \ref{eq:stoch} and \ref{eq:combine}, using 50\% overlapping and Hann-windowed segments of duration $T=4$ sec, 0.25 Hz frequency resolution over the 7-200 Hz band, and an $8^{\rm th}$ order high-pass filter with 3 dB frequency of 5 Hz. The noise PSDs used in this calculation are those used to generate the mock data, i.e. they are not estimated from the data as is usually done in SGWB searches due to the fact that the CBC signals overwhelm the noise contribution to the strain power. 

\begin{figure*}[!htbp]
\includegraphics[width=0.95\textwidth]{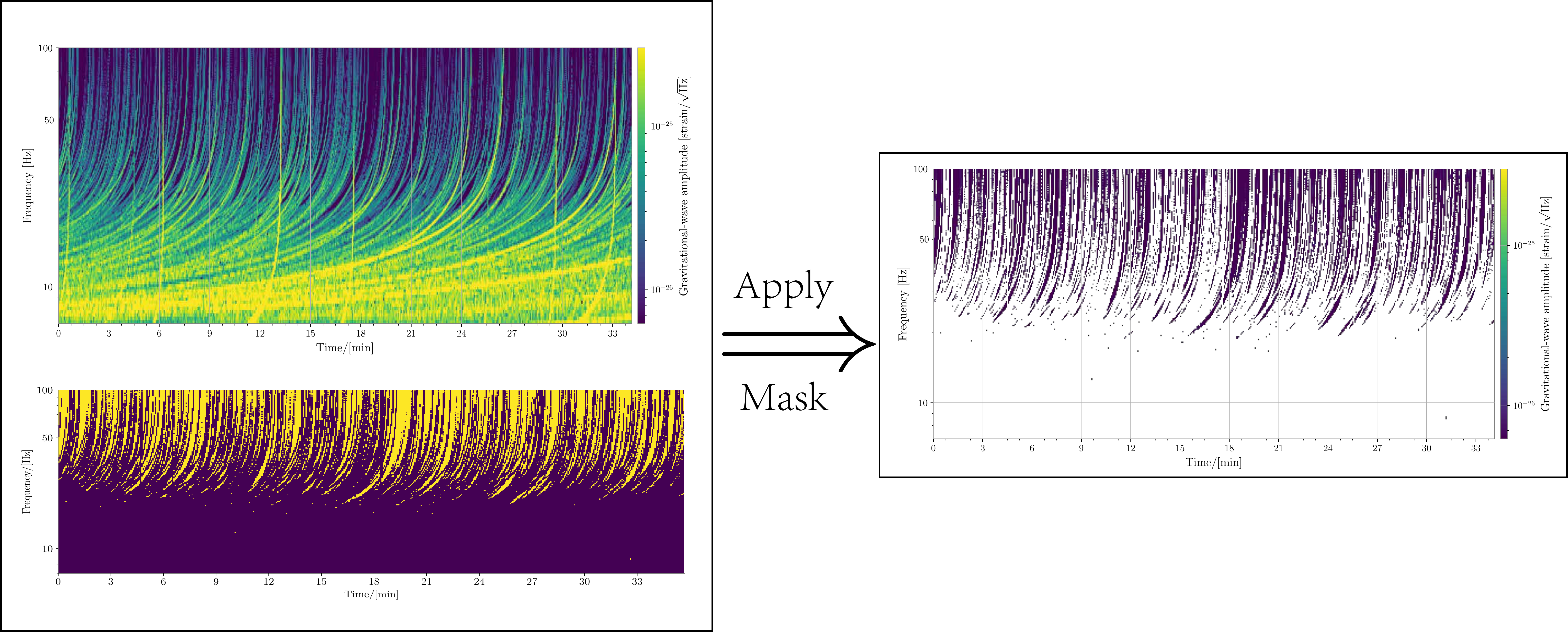}
\caption{An example of a spectrogram simulated for one of the detectors is shown in the upper left. The lower left panel shows the time-frequency mask capturing the CBC signals featured in this data segment (yellow denotes time-frequency pixels that are not masked). Applying this mask to the upper-left spectrogram yields the residual spectrogram shown on the right. We note that this figure is only illustrative---the masking in our study is done in cross-correlation spectrograms and not in individual detector spectrograms. Note that most of the pixels below 25 Hz are removed by the mask. }
\label{fig:example}
\end{figure*}

Computing the $\hat{C}_{\text{IJ}}$ statistic over all available time segments and frequencies (c.f. Eq. \ref{eq:combine}) would include all of the CBC foreground signals and necessarily result in a large $\Omega_\alpha$ estimate. Figure \ref{fig:example} (upper-left) shows an example of the strong CBC contribution to one detector's strain spectrogram. In order to remove the CBC foreground, we will attempt to mask all individual CBC signals. 

In a search using real data, each of these CBC signals would have to be individually detected and its parameters would have to be estimated. One would then estimate the GW chirp signal for each CBC event and mask the corresponding time-frequency pixels in the spectrogram. At the moment, such an analysis is not possible since the technology to simultaneously detect (and estimate parameters of) $\sim 10^6$ overlapping CBC signals is not available. 

We therefore adopt a different approach and set a  threshold $A_0$ on pixel amplitude: 
all time-frequency pixels in the spectrogram containing pure CBC events (i.e. without noise) with amplitude larger than $A_0$ are recorded. These pixels are then removed from the average in Eq. \ref{eq:combine} when analyzing the frames containing CBC events along with noise and/or SGWB. The value of $A_0$ is tuned for each of the three populations A, B, and C. In particular, setting the global threshold $A_0$ to be too high would not remove all (weak) CBC signals resulting in contamination of the $\hat{C}_{\text{IJ}}$ statistic. On the other hand, $A_0$ should be sufficiently high to avoid notching pixels due to the spectral leakage from the CBC signals---this would result in unnecessary removal of pixels, and therefore reduce the sensitivity of the SGWB search. The optimal choice of $A_0$ will therefore reduce the CBC contribution so as to make it smaller than the noise-only $\hat{\sigma}_{\text{IJ}}$ estimate.

Figure \ref{fig:example} (lower-left) shows an example of a mask computed using this procedure. When the mask is applied to the strain spectrogram of a single detector (upper-left), we obtain the residual spectrogram shown on the right panel. Note that the masking is particularly severe at low frequencies, where the many BNS signals overlap. In particular, frequencies below 25 Hz are effectively removed from the analysis. This has a strong impact on the sensitivity of the SGWB search since the lowest frequencies tend to be the most sensitive ones (note the $f^3$ term in Eq. \ref{eq:stoch}). We note that while Figure \ref{fig:example} is illustrating the strain spectrogram of one detector, in our analysis the masking is done on the cross-correlation estimator $\hat{C}_{\text{IJ}}(t_i;f_j)$.

We also note that the power from a given CBC signal could be shared between neighboring time segments, simply because the CBC signal may pass through multiple 4-sec long segments. Furthermore, since we are performing finite-time Fourier transforms, there could also be leakage of power into neighboring frequency bins. For both of these reasons, the track of a CBC signal in the spectrogram may be more than 1-pixel wide. Regardless, the masking procedure described above is capable of removing pixels containing such leaked power. 

On the other hand, the masking procedure defined above is too optimistic: while most of the BBH events in the universe may be detectable by CE detectors, a sizeable fraction of BNS events, especially at high redshifts, will not be \cite{TaniaCE}. To assess the importance of this unresolvable component of the CBC population, we estimate the energy density contained in CBC events with the CBC signal-to-noise ratio $\rho_{\rm CBC} < 8$, where
\begin{equation}
    \rho_{\rm I}=\sqrt{4\int_0^{\infty}\frac{|\tilde{h}_{\rm I}(f)|^2}{P_{n_{\rm I}}(f)}\D f},\quad \rho_{\rm CBC}=\sqrt{\rho_{\rm H}^2+\rho_{\rm L}^2},
\end{equation}
$\tilde{h}_\text{I}(f)$ is the frequency-domain CBC waveform, and $P_{n_{\text{I}}}(f)$ is the one-sided PSD of detector $\text{I}$.
We estimate $\rho_{\rm CBC}$ for each CBC event using the Python package \texttt{gwbench} \cite{gwbench}. 

The energy density of this unresolvable component of the CBC population may limit the sensitivity of CE detectors to the cosmological SGWB: individually unresolved CBC signals cannot be identified and therefore cannot be notched or subtracted from the real data. Furthermore, parameter estimation for each individually resolved CBC event will be associated with experimental errors, which in turn may lead to imperfect masking and to a residual CBC signal leaking into unmasked pixels. For these reasons, the masking procedure adopted here (based on the pixel amplitude threshold $A_0$) can be considered as optimistic---that is, this masking procedure will be more effective at removing the CBC signals than what will be possible in the analysis of real CE data.

\begin{table*}[htbp] 
\setlength{\abovecaptionskip}{0.05cm} 
\centering
\setcounter{table}{0}
\renewcommand{\tablename}{Table}
\begin{tabular*}{\hsize}{@{\extracolsep{\fill}}l|l|c|c|c}
  \hline
  \hline
  Population  \&  Interval&Case & $\hat{C}_{\rm IJ}$ & $\hat{\sigma}_{\rm IJ}$ & SNR\\
  \hline 
 \multirow{2}{*}{\diagbox{\ \quad\quad\quad\quad\quad\ \ \ \ \ }{}}&Noise+SGWB[$\alpha=0$]&4.0$\times10^{-11}$&9.0$\times 10^{-14}$&\textbf{4.5}$\bm{\times 10^2}$\\
  &Noise+SGWB[$\alpha=4$]&4.0$\times10^{-11}$&1.5$\times10^{-13}$&\textbf{2.6}$\bm{\times10^2}$\\
  \hline
  \hline 
  &Noise+CBC[$\alpha=0$]&9.3$\times 10^{-10}$&9.0$\times 10^{-14}$&1.0$\times 10^4$\\
  &Noise+CBC(removed)[$\alpha=0$] &3.2$\times 10^{-11}$&$5.1\times 10^{-11}$ &0.63\\
  Pop A:&Noise+CBC(removed)+SGWB[$\alpha=0$] &$7.2\times 10^{-11}$&$5.1\times 10^{-11}$&\textbf{1.4}\\
  \multirow{2}{*}{BNS=98.0\%,BBH=2.0\%; $d$=3s}&Noise+CBC(removed)+SGWB[$\alpha=4$]&3.9$\times10^{-11}$&1.1$\times10^{-12}$&\textbf{35}\\
  &Noise+CBC($\rho\leqslant8$)[$\alpha=0$]&$\sim 8.6\times 10^{-11}$&-&-\\
&Noise+CBC($\rho\leqslant8$)[$\alpha=4$]&$\sim 5.6\times 10^{-11}$&-&-\\
\hline
\hline
 &Noise+CBC[$\alpha=0$]&6.5$\times 10^{-10}$&9.0$\times 10^{-14}$&7.2$\times 10^3$\\
  Pop B:&Noise+CBC(removed)[$\alpha=0$] &1.9$\times 10^{-12}$&$2.7\times 10^{-12}$&0.68\\
  \multirow{2}{*}{BNS=95.9\%,BBH=4.1\%; $d$=6s}&
  Noise+CBC(removed)+SGWB[$\alpha=0$]&$4.2\times 10^{-11}$&$2.7\times 10^{-12}$&\textbf{15}\\
  &Noise+CBC(removed)+SGWB[$\alpha=4$]&$3.9\times 10^{-11}$&$3.5\times 10^{-13}$&$\bm{1.1\times10^{2}}$\\
  &Noise+CBC($\rho\leqslant8$)[$\alpha=0$]&$\sim 4.2\times 10^{-11}$&-&-\\
  &Noise+CBC($\rho\leqslant8$)[$\alpha=4$]&$\sim 2.8\times 10^{-11}$&-&-\\
\hline
\hline
  &Noise+CBC[$\alpha=0$]&4.1$\times 10^{-10}$&9.0$\times 10^{-14}$&4.6$\times10^3$\\
  &Noise+CBC(removed)[$\alpha=0$]&7.2$\times 10^{-13}$&1.3$\times 10^{-12}$&0.56\\
 Pop C: &Noise+CBC(removed)+SGWB[$\alpha=0$]&4.1$\times10^{-11}$&
  1.3$\times 10^{-12}$& \textbf{31}\\
  \multirow{2}{*}{BNS=83.3\%,BBH=16.7\%; $d$=24s}&Noise+CBC(removed)+SGWB[$\alpha=4$]&4.0$\times10^{-11}$&
  2.4$\times 10^{-13}$& $\bm{1.7\times10^{2}}$\\
  &Noise+CBC($\rho\leqslant8$)[$\alpha=0$]&9.1$\times 10^{-12}$&-&-\\
  &Noise+CBC($\rho\leqslant8$)[$\alpha=4$]&6.0$\times 10^{-12}$&-&-\\
  \hline
  \hline
\end{tabular*}
\caption{Results of the notching algorithm are shown for three population models A, B, and C. For each model we show the SGWB point estimate $\hat{C}_{\rm IJ}$, the corresponding standard deviation $\hat\sigma_{\rm IJ}$, and the signal-to-noise-ratio ${\rm SNR} = \hat{C}_{\rm IJ}/\hat\sigma_{\rm IJ}$. The corresponding thresholds $A_0$ for Pop A, B and C are equal to $8\times10^{-30}/\sqrt{\text{Hz}},5\times10^{-29}/\sqrt{\text{Hz}}$ and 5$\times10^{-30}/\sqrt{\text{Hz}}$, respectively. For each POP model and for two values of $\alpha=0,4$, several analyses are completed: (i) CBC-only simulation before any notching, (ii) CBC-only simulation after notching assuming all CBC signals are resolved, (iii) CBC+SGWB simulation after notching assuming all CBC signals are resolved, and (iv) CBC-only simulation without notching in which only CBC signals with $\rho_{\rm CBC} \le 8$ are included. See text for further discussion.}
\label{table:1}
\end{table*}

\textbf{Results:} Table \ref{table:1} summarizes the results of the notching algorithm discussed above. For each of the three populations (A, B, and C) 
and for two values of $\alpha$ (0 and 4)
we perform several analyses. Starting with $\alpha=0$, we first simulate the GW time series of the CBC population using realistic CE noise realization and we estimate the SGWB statistic $\hat{C}_{\rm IJ}$ and the corresponding standard deviation $\hat\sigma_{\rm IJ}$ without any notching. This results in a very large SGWB signal-to-noise ratio ($\hat{C}_{\rm IJ}/\hat\sigma_{\rm IJ}$), ranging from 4600 for Pop C which has the fewest BNS events to 10,000 for Pop A which has the most BNS events. 

Second, we determine the optimal pixel amplitude threshold $A_0$ for each population, apply it to notch out the CBC signals, and obtain the SGWB statistic after notching (for $\alpha=0$). For all three Pops our procedure results in $\hat{C}_{\rm IJ}/\hat\sigma_{\rm IJ}< 1$, indicating that we successfully removed the CBC population from the simulated data. The resulting SGWB standard deviation $\hat\sigma_{\rm IJ}$ ranges between $1.3 \times 10^{-12}$ for Pop C and $5.1 \times 10^{-11}$ for Pop A, revealing that the notching procedure leaves significantly different amounts of un-notched pixels in the three population cases. 

\begin{figure}
    \centering
    \includegraphics[width=0.48\textwidth]{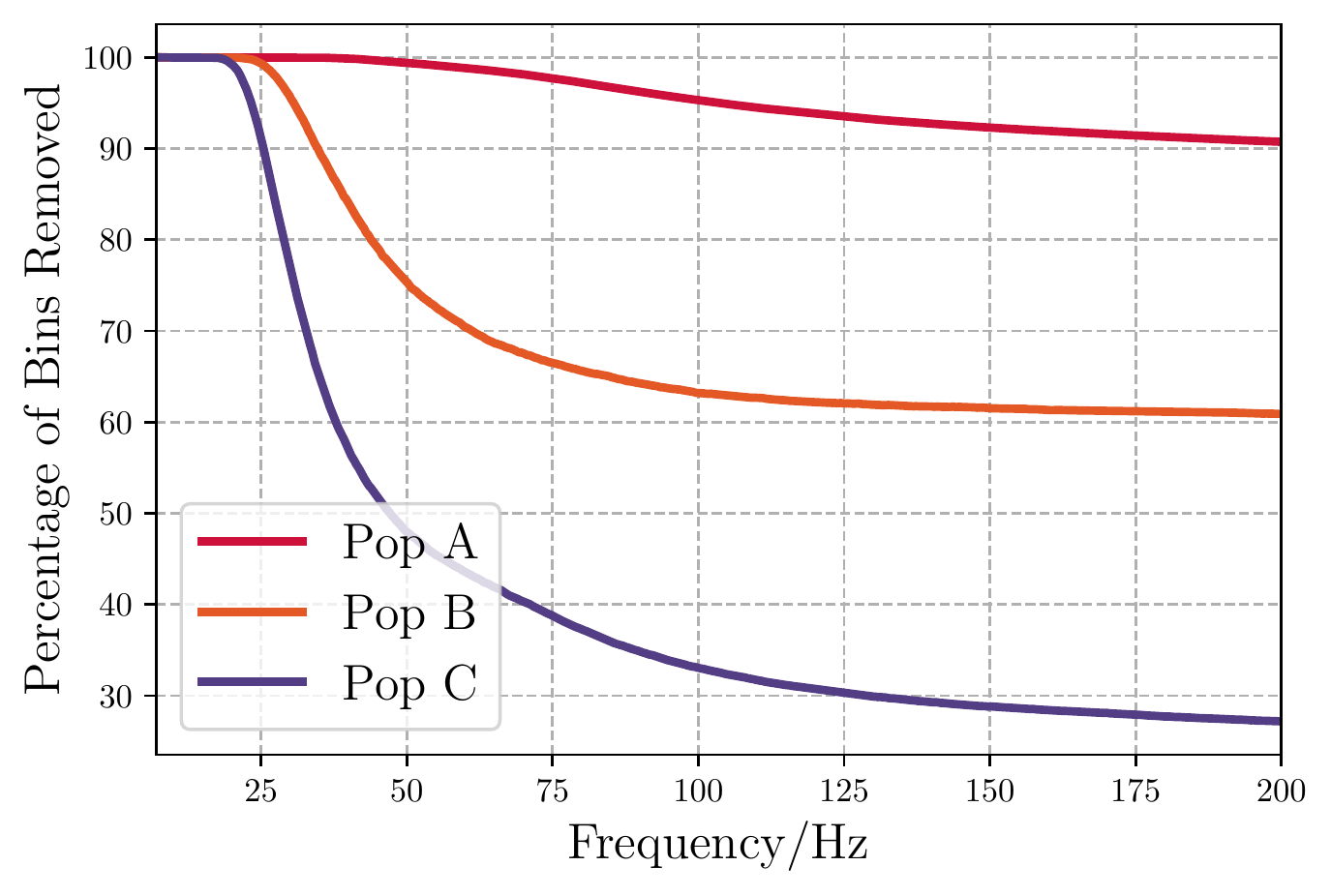}
    \caption{Percentage of time-frequency pixels removed by notching is shown as a function of frequency for all three population models. Below 25 Hz nearly 100\% percent of the pixels are removed for each population model. Above 25 Hz, the removed pixel percentage strongly depends on the population model. }
    \label{fig:3}
\end{figure}

Fig. \ref{fig:3} shows the fraction of removed pixels as a function of frequency. For all three populations, nearly all pixels below 25 Hz are notched out. Above 25 Hz there are significant differences among the three population models: for Pop C only $\sim 30\%$ of the pixels are removed in the region of 100-200 Hz, while for Pop A more than 90\% of pixels are notched out. This is consistent with the projected sensitivities $\hat\sigma_{\rm IJ}$ discussed above.

We also note that the optimal threshold $A_0$ values are significantly different among the three populations. For Pop C (fewest BNS events), the optimal threshold has the lowest value of $A_0 = 5\times 10^{-30}/\sqrt{\text{Hz}}$, for Pop B (medium number of BNS events) the corresponding threshold has the highest value of  $A_0 = 5\times10^{-29}/\sqrt{\text{Hz}}$, while for Pop A (largest number of BNS events) the threshold decreases again to $A_0 = 8\times 10^{-30}/\sqrt{\rm Hz}$. This pattern comes from the difficulty in dealing with CBC power leakage into the nearby pixels. For Pop C, successive events do not overlap as much in the low frequency region (where CBC tracks are nearly horizontal) as for Pop A and B, so a lower $A_0$ value is needed to notch out pixels with leaked power. For Pop B there are $\sim 4 \times$ more BNS events as compared to Pop C, so the low-frequency overlap is stronger and a weaker (larger) $A_0$ value suffices. In both cases, nearly all pixels below 25 Hz are removed. For Pop A, which has $\sim 8 \times$ more BNS events than Pop C, the leakage in the high frequency region becomes more significant, and therefore requires a lower value of $A_0$. Again, this is consistent with the pattern shown in Fig. \ref{fig:3}.  

As the third analysis (for each population, $\alpha=0$), we add a truly stochastic component to the simulated data with energy density $\Omega_0=4\times 10^{-11}$. As shown in Table \ref{table:1}, this SGWB is successfully recovered in Pop B and C simulations, which are sufficiently sensitive after notching, but not in the Pop A simulation where the SGWB standard deviation after notching is $\hat\sigma_{\rm IJ} = 5.1 \times 10^{-11}$, i.e. larger than the injected SGWB. Of course, in all three cases the simulated SGWB is completely dwarfed by the CBC population before notching, and is therefore undetectable. 

As the final analysis for $\alpha = 0$, for each population we estimate the SGWB statistic and standard deviation by adding up the CBC signals with CBC signal-to-noise $\rho_{\rm CBC} < 8$. This estimate illustrates the unresolved part of the CBC population, which will not be possible to individually notch or subtract in the analysis of real CE data. In all three cases, the unresolvable CBC SGWB estimator $\hat{C}_{\rm IJ}$ is larger than the standard deviation after notching, ranging from $9.1 \times 10^{-12}$ for Pop C to $8.6 \times 10^{-11}$ for Pop A. This result implies that the sensitivity of the SGWB searches with real CE data will not be limited by the ability to notch out resolvable CBC signals, but by the residual energy density in the unresolved part of of the CBC population. We note that the numerical computation of this unresolvable foreground requires large computational resources---consequently, we performed the full calculation only for Pop C and then scaled it to get estimations for Pops A and B. The scaling was performed based on the relative numbers of BNS events in the three population models, since the CBC parameter distributions were otherwise the same in the three models and the BBH population is almost entirely resolvable. 

We also repeat the above analyses for $\alpha=4$ case, which favors higher frequencies and is therefore less susceptible to the heavy notching at frequencies below 25 Hz. Consequently, 
in the simulations of the CBC foreground combined with the blue SGWB ($\alpha=4$, $\Omega_4 = 4.0 \times 10^{-11}$) after notching we find that $\hat{\sigma}_{\rm IJ}$ is $\sim 10\times$ smaller than the values we obtained for $\alpha=0$, for each of the three population cases. The simulated SGWB is now recovered well for all three population cases (with large SNR), indicating that the loss of low frequencies has a much smaller impact on blue SGWB energy density searches. We have also computed the point estimate $\hat{C}_{\rm IJ}$ for the unresolved part of the CBC population assuming $\alpha=4$ in the search, and similarly to the $\alpha=0$ case we find that this point estimate is larger than $\hat\sigma_{\rm IJ}$ obtained after notching (for each of the three population models). Therefore our conclusion carries over: the search for $\alpha=4$ SGWB will also not be limited by the ability to notch the individually resolved CBCs, but rather by the unresolved CBC contributions. However, the unresolved CBC limit now takes smaller values, ranging between $6.0\times 10^{-12}$ for Pop C to $5.6\times10^{-11}$ for Pop A.

\textbf{Conclusions:} The primary aim of this paper is to investigate whether the currently available technology of notching the CBC signals in time-frequency space is sufficient for the CBC foreground removal in SGWB searches with the third-generation gravitational wave detector data. The answer, based on the presented results, appears to be ‘yes’: for each of the three CBC population models and for two spectral indices ($\alpha = 0,4$), the notching procedure implemented here was more than sufficient to reach the SGWB sensitivity floor defined by the unresolvable component of the CBC population. This floor is roughly on the scale $\Omega_{\rm GW} \sim 10^{-11}$ and it depends on the CBC population model and on the spectral index of the SGWB energy density. In particular, the floor is lower for populations having fewer CBCs and for larger values of $\alpha$.

We highlight, however, some limitations of our analysis that should be investigated in future works. First, we did not add noise glitches in our simulated data. Such glitches may impact both the CBC estimation and the SGWB search, so future studies should explore how significant they may be. Second, the notching we performed is based on {\it known} CBC parameters used in the simulation and not on the {\it estimated} ones that will be available in the analysis of real CE data. While currently the technology for estimating parameters of $>10^5$ overlapping CBC signals does not exist, it should be possible to include representative CBC parameter uncertainties in the notching procedure to study the impact of the imperfect CBC signal removal. Third, our analysis relied on simulated detector noise PSDs which will not be available in the analysis of real CE data. Possible approaches to handle this challenge include estimating the PSDs from real data after the CBC signals are removed, estimating the detector noise PSDs from first principles (i.e. from the known noise sources), and modeling and estimating the noise PSDs simultaneously with the SGWB. We also note that in the case of the Einstein Telescope detectors, it will be possible to define the null stream that can provide an unbiased estimate of the noise PSD~\cite{ETnull}. Fourth, the standard time-frequency Fourier decomposition with uniform pixel size might not be optimal for the notching procedure. For example, wavelet decompositions that allow extended pixels in time (i.e. horizontal rectangular pixels) at low frequencies and extended pixels in frequency (i.e. vertical rectangular pixels) at high frequencies may be better suited to minimize power leakage. And fifth, techniques have been proposed to subtract CBC signals \cite{TaniaCE,CutlerHarms,ShwarmaHarms} or fit them simultaneously with the SGWB \cite{Thrane_TBS,PhysRevLett.125.241101}, rather than to notch them out. These approaches would preserve more data and therefore result in smaller $\hat\sigma_{\rm IJ}$, but they might still be limited by the sensitivity floor due to unresolved compact binaries.\cite{zhoubei1,zhoubei2}


\textit{Acknowledgements:}
The authors are grateful for computational resources provided by the LIGO Laboratory and supported by National Science Foundation (NSF) Grants PHY-0757058 and PHY-0823459. The work of VM was supported by the NSF grant PHY-2110238.

\bibliography{apssamp}

\end{document}